# Molecular dynamics simulation of high slip flow of water confined between graphene nanochannels at experimentally accessible strain rates

Carmelo Riccardo Civello[1], Luca Maffioli[1],

Edward R. Smith[2], James P. Ewen[3,4], Peter J. Daivis[5],

Daniele Dini[3], B. D. Todd[1]

[1]Department of Mathematics, Faculty of Science, Computing and Emerging Technologies, Swinburne University of Technology, PO Box 218, Hawthorn, Victoria 3122, Australia

[2]Mechanical and Aerospace Engineering, Brunel University London, Kingston Lane, Uxbridge, UB8 3PH, London, United Kingdom

[3]Department of Mechanical Engineering, Imperial College London, South Kensington Campus, London, SW7 2AZ, London, United Kingdom

[4]Department of Mechanical Engineering, University of Bath, Bath BA2 7AY, U.K

[5]Department of Physics, RMIT University, GPO Box 2476, Melbourne, 3001, Victoria, Australia

**Corresponding author:** B. D. Todd, `btodd@swin.edu.au`

March 17, 2026

## Abstract

The transient time correlation function method (TTCF) has emerged as a powerful methodology for accurately probing systems at low shear rates. In the present study, TTCF was used to evaluate the shear rate dependence of the slip length in a high-slip system consisting of water confined between graphene walls at experimentally accessible shear rates, for which classical nonequilibrium molecular dynamics (NEMD) is unfeasible. The corresponding Navier friction coefficient was computed for all shear rates spanning six orders of magnitude and compared with the equilibrium limit. We report for the first time NEMD results obtained at experimentally accessible shear rates using the TTCF approach for a system that has attracted significant interest over the past decades. The slip length calculated with TTCF is in good agreement with previous equilibrium molecular dynamics simulations and experiments. Our aim here is to highlight the extraordinary power of TTCF, particularly for high-slip (low strain-rate) systems, and to verify that equilibrium methods directly match NEMD measurements at experimentally accessible strain rates.

# 1  Introduction

Nonequilibrium molecular dynamics (NEMD) has proven to be highly effective for probing systems far from equilibrium, where strong driving fields generate clear responses that can be characterised with good statistical accuracy. However, a well-known limitation of NEMD is its inability to capture the response of systems subjected to weak external fields, for which the signal-to-noise ratio (SNR) vanishes. Since such "weak" fields are the physically relevant ones, direct NEMD simulations can only be used to extrapolate the system's response to the weak-field limit. As a result, and despite various attempts to resolve such limitations in the past decades [1, 2], its capability to describe the tribology or rheology of complex fluids, as well as experimental systems undergoing realistic strain rates, is heavily constrained.

In the 1980s, Evans and Morris derived the so called Transient-Time Correlation Function method (TTCF) for the thermostatted nonlinear response to planar Couette flow [3–5]. They provided an optimised implementation and proved its validity and agreement with standard NEMD for simple fluids. In the equilibrium limit, the TTCF and Green–Kubo [6] methods are equivalent, making the TTCF approach as efficient as the Green–Kubo formulation, with the latter being restricted to linear response [4]. As a consequence, the TTCF method is capable of bridging the linear and non-linear response, making it an extremely powerful molecular simulation tool to model realistic strain rates for complex fluids.

Since these early studies, the TTCF methodology has been used in a number of applications [7–23] including the study of confined systems at experimentally accessible shear rates. Currently, we are aware of only three studies that involve the use of TTCF to investigate molecular fluids [11, 16, 21]. It is also worth mentioning that Maffioli et al. [23] have recently developed a tool that enables users to easily access and use the TTCF algorithm incorporated within it and which is readily interfaced with the LAMMPS software package [24].

In the present work, we apply the TTCF method to water confined between graphene walls, focusing on the computation of slip length and the Navier friction coefficient. We then compare our TTCF results with the equilibrium method developed by Varghese et al. [25], which in turn is based on the technique developed by Hansen et al. [26]. As a simple fluid, water does not strictly require the use of TTCF, as its linear response regime can be investigated with direct NEMD at sufficiently high shear rates. Neverthe-less, systems involving water and graphene or carbon nanotubes have attracted considerable interest over the past decades, with their exceptionally high slip representing one of the most distinctive interfacial properties [27–30]. Moreover, direct NEMD measurements are characterized by large statistical uncertainty for high slip systems [31]. Given the limited number of studies employing TTCF to investigate molecular fluids, and the aforementioned interest in high-slip systems, we explore the applicability of TTCF to this type of system. Our results provide insight into both the reliability of TTCF for molecular fluids and the interfacial friction characterising water–carbon systems, which remain highly relevant for nanofluidic applications. To our knowledge, this is the first study in which the water-graphene system is studied at experimentally accessible strain rates via NEMD simulation, allowing us to make direct comparisons to comparable experimental measurements.

## 2 Methodology

### 2.1 Transient-Time Correlation Function

Several authors in the past have derived the nonlinear response for a generic phase variable of systems exposed to an external field [5, 32–34]. This response is referred to as the Transient-Time Correlation Function (TTCF), and its most general expression for an arbitrary physical quantity $B$ is as follows:

$$\langle B(t) \rangle = \langle B(0) \rangle + \int_0^t ds \langle B(s) \Omega(0) \rangle. \tag{1}$$

Here, the $\langle \cdot \rangle$ operator denotes the ensemble average and its definition is given in Eq. (2). $\Omega$ represents the dissipation function, proportional to the rate of energy dissipation and defined as $\Omega = -\dot{U}/k_b T$, where $U$ is the total energy of the system. It is important to note that the external field is turned on at $t = 0$ while the system is in equilibrium. This formulation establishes a connection between the steady state of a generic phase variable and the energy dissipation of the initial state. Specifically, the steady state depends on its correlation with $\Omega(0)$. The convergence of the TTCF integral is ensured when the system exhibits mixing. Two quantities are said to mix when they become statistically uncorrelated after a long enough time, and so when $t \to \infty$, $\langle B(s)\Omega(0) \rangle \to \langle B(s) \rangle \langle \Omega(0) \rangle$. For a detailed discussion, the reader is referred to [34].

Formally, the ensemble average, typically the standard approach in NEMD, involves an integral over phase space weighted by the ensemble probability density. In practice, molecular dynamics provides a discrete sampling of configurations drawn from this distribution, allowing the ensemble average to be approximated by a simple arithmetic mean over the *N* sampled configurations:

$$\langle B(t) \rangle = \frac{1}{N} \sum_{i=1}^{N} B_i(t). \qquad (2)$$

Equations (1) and (2) are formally equivalent and yield the same result. The difference lies in the fact that the TTCF formulation is based on the correlation with the dissipation function at $t = 0$, whereas Eq. (2) involves a direct evaluation of the ensemble average at each time. For this reason, we refer to Eq. (2) as the direct average (DAV), following the terminology previously adopted by other authors [22]. Owing to this distinction, the two methods exhibit different practical performance, as will be shown later in this work. It is important to stress that the data used for the two approaches are the same and they are obtained from the same set of simulations; what changes is the way they are processed.

## 2.2 Molecular system setup

One drawback of the TTCF method is its significant computational cost, which is necessary to obtain high-quality results. It requires a substantial number of independent trajectories, as observed and commented by various authors and noted specifically in specialised text books [6, 32, 34]. Furthermore, molecular system simulations are inherently more computationally demanding than simpler cases like the Lennard-Jones fluid. For these reasons, we aim to strike a balance between computational efficiency and the physical reliability of the results.

Accordingly, the extended simple point charge (SPC/E) model is employed to describe the water molecules [35], owing to its relatively low computational cost and reliability compared to more complex water models. The SHAKE algorithm [36] is used to constrain the molecule, while the Tersoff three-body potential is employed for the graphene sheets [37, 38]. The chosen models are the same ones utilised by Varghese et al. [25]. Both of these models have proven suitable for the intended purpose. A Lennard-Jones short-range Coulombic pair potential with a cutoff of $r_c = 10$Å is used to describe the interactions between the water molecules and the graphene layers, as well as the interactions between the graphene layers themselves and water-water interactions. The Lennard-Jones parameters for the water-graphene interactions are taken from Werder et al. [39]. The particle–particle particle–mesh (PPPM) solver was used to compute the long-range Coulombic interactions with an accuracy of $10^{-5}$ [40].

A system with 320 water molecules was set up for this study. The water molecules were generated using Packmol [41] while the graphene sheets were generated in VMD using the Carbon/Boron Nitride Nanostructure Builder plugin. Subsequently, VMD was employed to merge the solid and liquid phases [42]. The walls are made of three A-B-A stacked graphene layers with a distance of 0.34 nm between the adjacent layers [43]. A recent study investigating the slip length of water on graphene substrates with varying numbers of layers showed that the slip length is nearly fully converged from three graphene layers onward [44].

The system's temperature was maintained at 300K using a Langevin thermostat [45] applied to the two inner layers of each wall, while a third external layer was kept fixed to maintain a constant volume. Although the TTCF algorithm is formally valid for deterministic systems, our simulations show that a Langevin thermostat, despite being stochastic rather than deterministic, still provides valid results for TTCF computations. We used this rather than a deterministic thermostat, such as a Nose-Hoover thermostat, for computational efficiency. The graphene layers had dimensions of approximately 22Å in both non-confining directions $x$ and $y$. Prior to fixing the channel width, the system pressure was equilibrated to 1 atmosphere by applying a force to the top wall while holding the bottom wall fixed, effectively mimicking a piston-like setup. The wall-to-wall channel width is $h \approx 2$ nm. In Ref. [46], the friction coefficient was shown to be independent of the channel width, whereas Refs. [47, 48] demonstrated, using ReaxFF [49] and the SPC/E model, respectively, that water exhibits enhanced viscosity and oscillatory behavior when $h < 2$ nm. Although such confinement-induced viscosity enhancement may therefore play a role in the present system, we consider the case $h \approx 2$ nm to be particularly relevant, as several previous computational studies have focused on comparable channel widths [50–52], and recent experimental investigations have explored water transport in channels with $h < 2$ nm [51, 53]. Also, the relatively low computational cost associated with simulating this system is not only a consequence of the reduced number of particles, but also of the narrow channel width, which leads to a faster convergence of the transient regime, as predicted by the analytical solution for Couette flow [54, 55]. A representation of the system is shown in Fig. S1 of the Supporting Material.

The system was then sheared by moving the walls with equal and opposite speeds ($\pm v$), taking advantage of the shorter transient than the case with only one sliding wall [55]. The value of $v$ was chosen such that the desired strain rate was achieved. Falk et al. [46] reported that the friction coef-ficient is independent of the graphene sheet orientation, whereas Wagemann et al. [52] demonstrated that orientation-dependent effects become appreciable only under strong external fields. On this basis, no preferential sliding direction is expected. Accordingly, in the present study the walls are displaced along the zigzag crystallographic direction. The use of a rigid model for the water molecule allows for a higher integration time-step to be used, thereby not having to deal with the high frequency oscillations of flexible harmonic bonds; hence, an integration time-step $\delta t = 2$ fs is used. Data are saved every 10 time-steps. Simulations were performed using the LAMMPS software package [24] and an in-house implementation of the TTCF algorithm was used.

## 2.3 TTCF algorithm

To run multiple nonequilibrium (*daughter*) trajectories an initial set of starting states is needed. This was generated from 8000 equilibrium (*mother*) independent trajectories [6, 32, 34]. For each mother, 125 states were sampled at a time interval of 3000 time-steps (6 ps). A time interval longer than the decay times of the velocity autocorrelation function and the stress autocorrelation function [56, 57] was chosen to ensure that each state was independent of the others. The daughter trajectories were followed for 10 ps. Each data point required about $2 \times 10^5$ cpu-hours. Periodically sampling the microstates from a trajectory is a straightforward method to produce a set of independent and identically distributed points whose exact statistical ensemble is generally unknown [22, 23]. To speed up the process, multiple independent equilibrium trajectories in the same thermodynamic state can be generated in parallel.

Since the convergence of the integral in Eq. (1) is guaranteed by the mixing nature of the system $(t \to \infty, \langle B(s)\Omega(0) \rangle \to \langle B(s) \rangle \langle \Omega(0) \rangle = 0)$ and by $\langle \Omega(t=0) \rangle = 0$, it is necessary to ensure that the empirical sample average of $\Omega(t=0)$ is also identically zero [4, 32]. This can be achieved by using the following phase space mappings

$$\begin{aligned}
(x_i, y_i, z_i, v_{xi}, v_{yi}, v_{zi}) &\to (x_i, y_i, z_i, v_{xi}, v_{yi}, v_{zi}), \\
(x_i, y_i, z_i, v_{xi}, v_{yi}, v_{zi}) &\to (x_i, y_i, z_i, -v_{xi}, -v_{yi}, -v_{zi}), \\
(x_i, y_i, z_i, v_{xi}, v_{yi}, v_{zi}) &\to (-x_i, y_i, z_i, -v_{xi}, v_{yi}, v_{zi}), \\
(x_i, y_i, z_i, v_{xi}, v_{yi}, v_{zi}) &\to (-x_i, y_i, z_i, v_{xi}, -v_{yi}, -v_{zi}).
\end{aligned} \quad (3)$$

The total number of simulated trajectories is four million. For each mapped trajectory, $\Omega(0)$ is first computed. After the four mapped trajectories have been simulated, the quantity $B(s)\Omega(0)$ is evaluated for each of them and subsequently used to calculate the ensemble average $\langle B(s)\Omega(0)\rangle$. This average is updated every time four new mapped trajectories are generated. The dissipation function $\Omega$ for the present system is derived in the Section S1 of the Supplementary Material. To perform statistical analysis, independent ensemble averages are obtained from the four million trajectories and details of the averaging procedure are provided in the following section. Each block average is then time-integrated according to Eq. (1) to compute the time correlation function, using the trapezoidal rule.

## 3 Results

### 3.1 TTCF validation

The first simulation aimed to verify the agreement between the direct averaged (DAV) NEMD quantities and the corresponding TTCF quantities, the accuracy of the algorithm, and ultimately the fine-tuning of the input parameters. Quantities derived via TTCF are characterised by a statistical error that grows linearly with time. This behaviour arises from the time integration of $\langle B(s)\Omega(0)\rangle$, whose fluctuations are time-independent. Therefore, to ensure an acceptable statistical uncertainty at steady state, a sufficiently large number of trajectories must be performed. We found that 5 million trajectories for each data point were satisfactory for our purposes. The results were then block-averaged to obtain 32 independent samples (each sample is obtained by averaging over $4 \times 10^6/32$ trajectories), on which we performed the statistical analysis. The shear rate applied to the system was $\dot{\gamma} = 5 \times 10^{10} \mathrm{s}^{-1}$, which is high enough to achieve a good signal-to-noise ratio (SNR) with the DAV which is taken as the benchmark in the present analysis.

As mentioned previously, the fluid is highly confined; therefore, its velocity profile near the walls deviates from the homogeneous continuum prediction. This is evident in Fig. 1, where we observe that the velocity profile far from the walls is approximately linear, while the first fluid layers adjacent to the walls deviate from it, taking values between the wall velocity and the linear prediction. As expected, the system exhibits high slip at the interface, with an almost flat velocity profile, consistent with previous studies [31]. Figure 1 also shows good agreement between the TTCF and DAV profiles within the statistical uncertainty. This is further highlighted by the velocity time evolution (Fig. 2) and the shear stress at the interfaces (Fig. S2). In particular, in the former case, the TTCF data are less affected by thermal noise, which in the DAV results appears as fluctuations around the average local velocity. In both cases, the velocity evolution exhibits an overshoot before settling into its steady state, consistent with previous results [55].

A weighted linear fit, with weights proportional to the inverse of the variance, was performed on both the DAV and TTCF velocity profiles (see Fig. S3), considering only the central region of the channel, which was approximately 9Å thick. The velocity profiles were obtained by time-averaging over the time interval $[3.0, 3.5]$ ps. The corresponding slip length was computed as the ratio between the slip velocity $v_s$, obtained from the linear fit, and the shear rate $L_s = v_s/\dot{\gamma}$. The slip length is equal to 45.4 nm and 49.5 nm (see Fig. 4 for the confidence intervals) for the DAV and TTCF measurements, respectively.

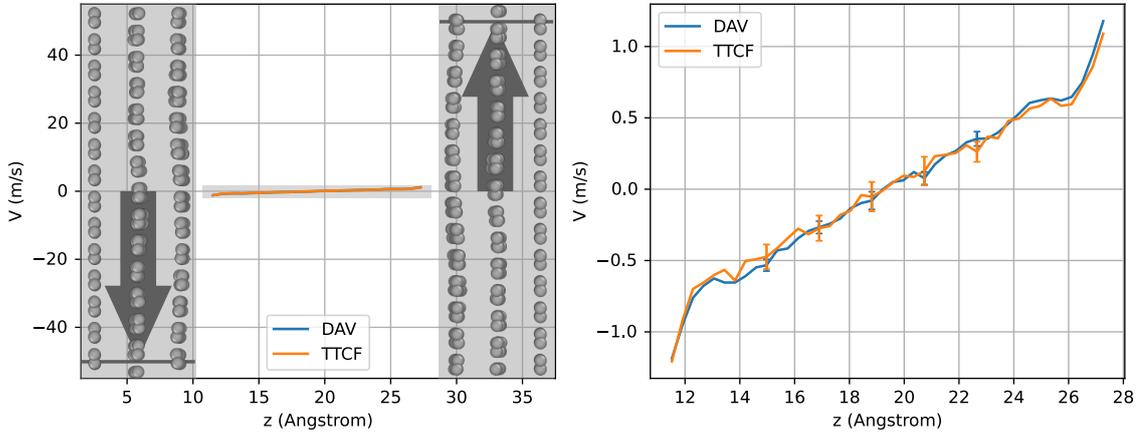

**Figure 1:** System velocity profile for both DAV and TTCF at $\dot{\gamma} = 5 \times 10^{10} \text{s}^{-1}$ ($v = 50$ m/s). In the left panel, the dark grey arrows and lines within the light grey regions (left and right), indicate the magnitude and direction of the wall velocity. The DAV velocity profile overlaps with the TTCF profile. The right panel shows a magnified view of the confined water velocity profile, corresponding to the light grey horizontal rectangle in the left panel. 95% confidence intervals are also shown for both DAV and TTCF data.

A simulation for a higher shear rate of $\dot{\gamma} = 5 \times 10^{11}$ s$^{-1}$ was also run. Related results are provided in the Supplementary Material (see Figures S4, S5).

## 3.2 Low shear regime

The shear rates we simulated span 6 orders of magnitude, ranging from $\dot{\gamma} = 5 \times 10^5$ s$^{-1}$ to $\dot{\gamma} = 5 \times 10^{11}$ s$^{-1}$. At lower shear rates, the DAV is not able to achieve a good SNR, and at any shear rates lower than $10^9$ s$^{-1}$ it would not be possible to even compare it with the TTCF. This is clearly shown for the velocity profiles in Fig. 3 (and for the shear stress in Fig. S6), for $\dot{\gamma} = 5 \times 10^7$. Fig. 4 shows the computed slip length for different shear rates. It is clear that for all the quantities computed by direct averaging, the SNR decreases, i.e., the error increases with decreasing shear rates, as we would expect. On the other hand, the opposite occurs for the TTCF calculations, consistent with trends observed in the study by Maffioli et al [23], i.e., the statistical error in the DAV is smaller than that of TTCF for very high but experimentally unrealistic strain rates, as seen for $\dot{\gamma} = 5 \times 10^{11}$ s$^{-1}$. These results once again confirm the remarkable superiority of TTCF to calculate all relevant physical properties of interest at strain rates that are experimentally accessible, while the usual DAV approach is futile.

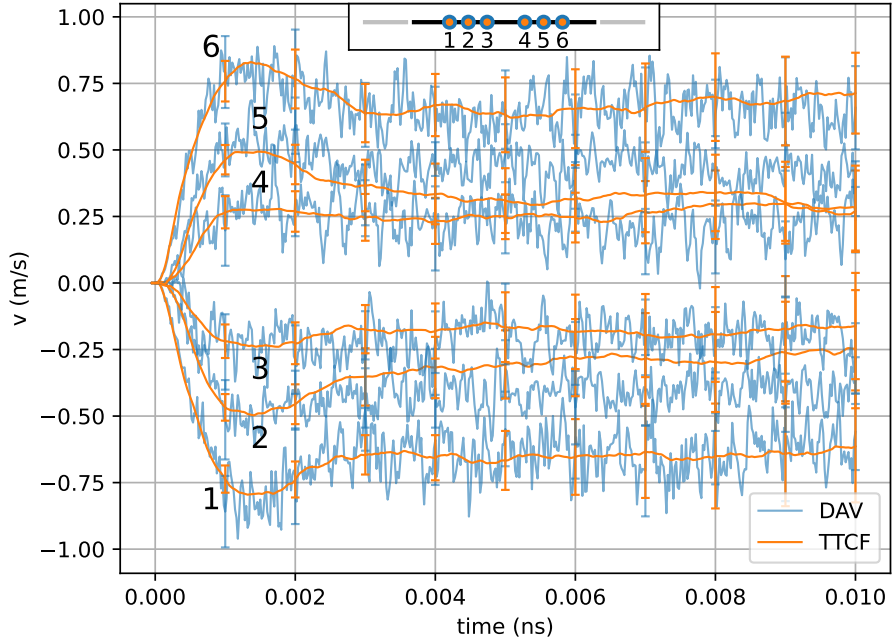

**Figure 2:** Water velocity evolution for both DAV and TTCF at coordinates $z = \{13.824, 15.744, 17.664, 21.504, 23.424, 25.344\}$ from curve 1 to curve 6, and for $\dot{\gamma} = 5 \times 10^{10} \mathrm{s}^{-1}$ ($v = 50$ m/s). 95% confidence intervals are shown for both DAV and TTCF data. The inset at the top represents the channel layout and the corresponding coordinates.

### 3.3 Comparison between NEMD and EMD simulations

We compare the low-shear Navier friction coefficient obtained from our NEMD simulations with that computed using the equilibrium method developed by Varghese et al. [25]. The NEMD Navier friction coefficient was calculated as

$$\xi = \frac{\tau_{zx}}{v_s},$$

where $\tau_{zx}$ is the shear stress at the fluid–solid interface, and $v_s$ is the slip velocity, both taken from the TTCF results. To evaluate $\tau_{zx}$ at the solid–liquid interface, the interaction force exchanged between the liquid phase and the designated wall was computed at each time step and divided by the wall surface area. $v_s$ was instead derived from the projection of the linear fit onto the interface plane. The EMD and TTCF results show good agreement within confidence intervals (Fig. 5). Clearly visible is the transition from linear to non-linear behaviour of the friction coefficient between $\dot{\gamma} = 5 \times 10^9 \text{s}^{-1}$ and $\dot{\gamma} = 5 \times 10^{10} \text{s}^{-1}$, where ξ decreases rapidly.

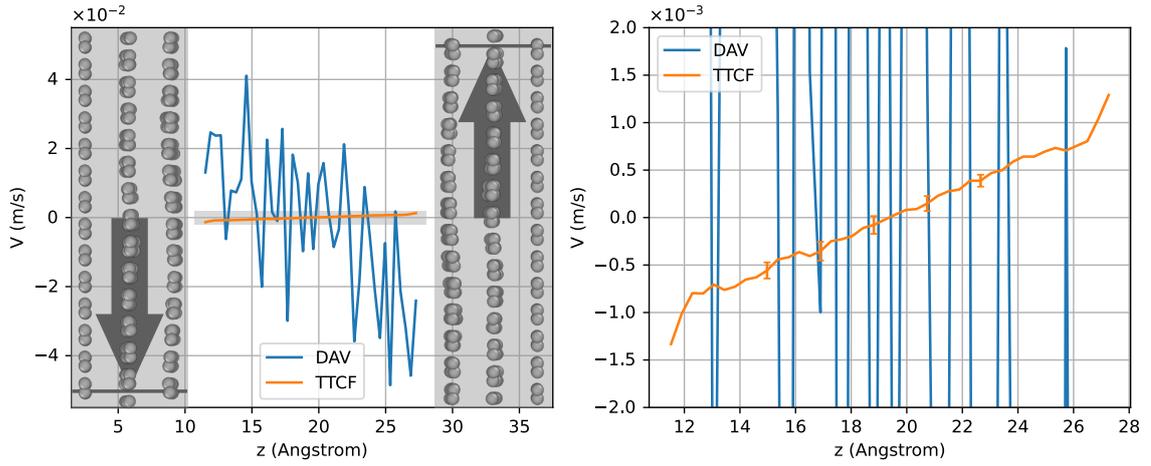

**Figure 3:** System velocity profile for both DAV and TTCF at $\dot{\gamma} = 5 \times 10^7 \text{s}^{-1}$ ($v = 0.05$ m/s) is shown in the figure. In the left panel, the dark grey arrows and lines within the light grey regions (left and right), indicate the magnitude and direction of the wall velocity. The right panel shows a magnified view of the confined water velocity profile, corresponding to the light grey horizontal rectangle in the left panel. 95% confidence intervals are shown only for the TTCF data.

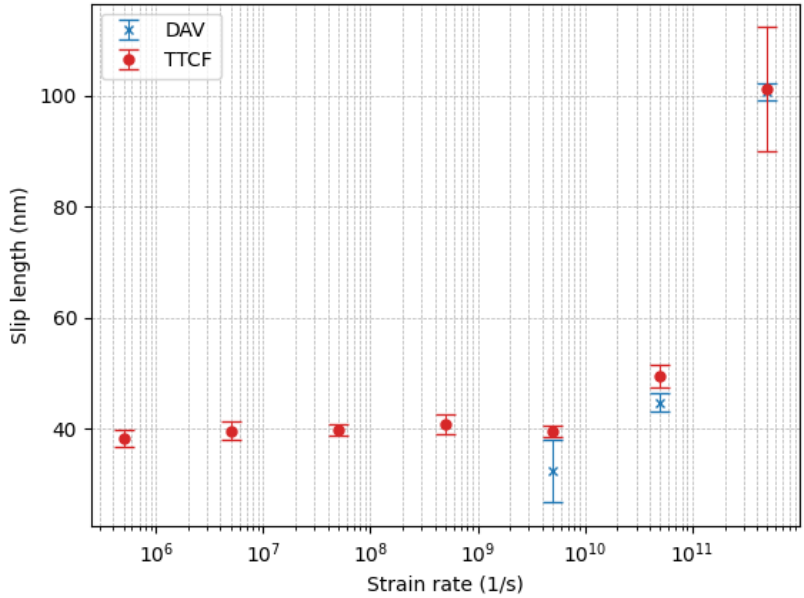

**Figure 4:** Slip length for different shear rates for both DAV and TTCF. DAV values for lower shear rates are not reported because they are out of range and characterized by very large errors.

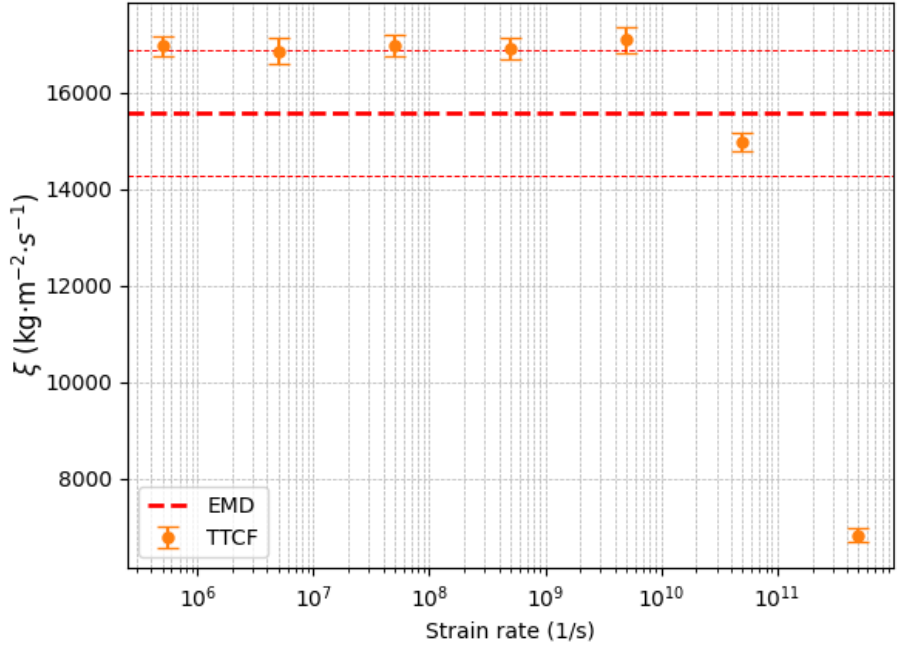

**Figure 5:** Navier friction coefficient for both the EMD and TTCF methods, the latter for different shear rates. The EMD coefficient (red dashed line) was computed using the equilibrium method derived by Varghese et al. [25]. Red dotted lines indicate the 95% confidence interval.

## 3.4 Comparison with the literature

It has previously been demonstrated that the friction coefficient of water flowing between carbon interfaces is sensitive to surface curvature [46]. For this reason, in the following discussion we compare our results with studies involving similar interfacial topologies. Comparisons with carbon nanotubes are therefore avoided. Molecular dynamics studies reported in the literature differ in several aspects, including the water model employed, the number of graphene layers composing the walls, and the surface topology. Moreover, different types of flow have been used, such as Poiseuille flow, Couette flow, and counterforce-driven flow. These differences make direct comparisons between studies somewhat problematic. The slip length is an intrinsic property of the system and should be independent of the type of flow considered. However, the flow type becomes relevant when comparing results obtained under different regimes, such as the linear or nonlinear response regimes. Therefore, we limit the comparison of our NEMD/TTCF slip length to the computational study of Kannam et al. [31], for which the shear rates used in their Couette flow simulations are reported. It should be noted that the SPC/Fw water model was employed in that work, and Celebi et al. [58] showed that this model yields larger slip lengths compared to the SPC/E model used in the present study. We also compare our result to the range of computational slip lengths summarised in [58] for the SPC/E model (see Fig. 6).

In the study by Kannam et al. [31], slip was investigated for a similar water–graphene system under Couette flow using direct NEMD as well as the original EMD method developed by Hansen et al. [26]. The main differences from the present work lie in the channel width ( $h \approx 4$ nm ), the water model employed (they used SPC/Fw [59]), and the integration time step ($\delta t = 1$ fs). To determine the slip length, the authors tested several approaches. Using the method where the error was evaluated through uncertainty propagation, they reported $L_s = 65 \pm 2$ nm for $\dot{\gamma} \sim 10^{10}$ s$^{-1}$ and $L_s = 113$ nm for $\dot{\gamma} \sim 10^{11}$ s$^{-1}$ (no uncertainty was provided for this value). In the present study, the slip length obtained with the TTCF method is $L_s = 49.5 \pm 2.1$ nm and $L_s = 102.7 \pm 10.9$ nm for $\dot{\gamma} \sim 10^{10}$ s$^{-1}$ and $\dot{\gamma} \sim 10^{11}$ s$^{-1}$, respectively. Celebi et al. [58] also compare the computational slip length obtained from different studies using different techniques, and water models. The slip lengths reported for the SPC/E model all range between 29 and 80 nm, confirming the agreement of the present study with the existing literature. Furthermore, in Ref. [50], a friction coefficient of $1.48 \times 10^4$ kg/(m$^2$s) was reported for a system with single-layered walls, in good agreement with our results ($\xi = 1.56 \times 10^4$ kg/(m$^2$s) for EMD and $\xi \approx 1.69 \times 10^4$ kg/(m$^2$s) for the NEMD linear regime). In Ref. [46], a friction coefficient in the range $1.40$ -$1.50 \times 10^4$ kg/(m$^2$s) was obtained for water confined between flat graphene sheets using the SPC/E model, whereas a larger value of approximately $2.05 \times 10^4$ kg/(m$^2$s) was obtained in Ref. [25]. In Fig. 6 we also compare the NEMD/TTCF friction coefficient characterizing the linear regime of the system with values reported in some of the works available in the literature related to the SPC/E water model.

Over the past decade, several experiments have investigated water flow over flat graphene surfaces using two-dimensional rectangular channels, in which two dimensions are orders of magnitude larger than the third (*h*). Among the earliest, Xie et al. [60] performed capillary flow experiments using a hybrid nanochannel design and reported a median slip length of 16 nm for *h* between 24 and 124 nm. This value exhibited a large scatter (0–200 nm) with no clear dependence on *h*, which was attributed to surface charges on graphene and interactions with the underlying silica substrate. Surface charges were argued to perturb the interfacial potential energy landscape, reducing both its smoothness and the effective hydrophobicity of graphene, while incomplete screening of the polar silica support by a single graphene layer may further suppress slippage. Evaporation-driven flow through Å-scale capillaries was later measured by Ashok Keerthi et al. [53] using microgravimetry for $h$ ranging from 0.68 Å to 8.5 nm, from which $L_s \approx 60$ nm was inferred. More recently, Kuan-Ting Chen et al. [61] employed a three-dimensional flow model accounting for non-uniform velocity profiles and performed measurements for $h = 45$ nm, obtaining $L_s = 40$ nm and $L_s = 33$ nm. They further showed that Refs. [53, 60] relying on two-dimensional models systematically overestimated the slip length. Also, Radha et al. [51] had previously measured evaporation driven flow through Å-scale capillaries but they did not derive the related slip length. In the literature, there is also another set of studies in which the slippage of water on graphene layers was measured via Atomic Force Microscopy (AFM) [62–64]. They all report similar slip lengths, $4.3 \pm 3.5$ nm [62], $6.8 \pm 2.9$ nm [63], and 8 nm [64], which are smaller than the values discussed so far. In particular, Li et al. [62] investigated slip length translucency by measuring slip for different numbers of graphene layers ($n$) and different underlying substrates. They highlighted a converging slip length for $n \geq 3$, independently of the substrate, at a shear rate of $10^3$-$10^4$ s$^{-1}$. They suggested that the smaller slip lengths obtained, compared to values reported in the literature, may arise from the different systems investigated, namely unconfined water. In Fig. 6 we also include the slip lengths obtained from the experimental studies discussed here. Although experimental measurements are sensitive to surface charge, roughness, and the nature of the supporting substrate, and computational studies employ different modelling approaches often disregarding the quantum contribution to the friction coefficient [65], the slip length obtained in the present work compares favourably with the results reported in most of the aforementioned studies.

# 4 Conclusions

In this work, we have demonstrated the applicability of TTCF for molecular fluids in high-slip systems and quantified the slip behavior of water flowing over flat graphene surfaces by means of non-equilibrium molecular dynamics simulations, extracting slip lengths over a range of applied shear rates. For the first time, the Transient Time Correlation Function approach was employed for this system to access experimentally relevant shear rates. In the linear response regime, the TTCF results were found to be consistent with the friction coefficient obtained from equilibrium molecular dynamics and in good agreement with previous computational studies. Furthermore, comparison with available experimental measurements of water flow over flat graphene surfaces revealed close quantitative agreement, supporting

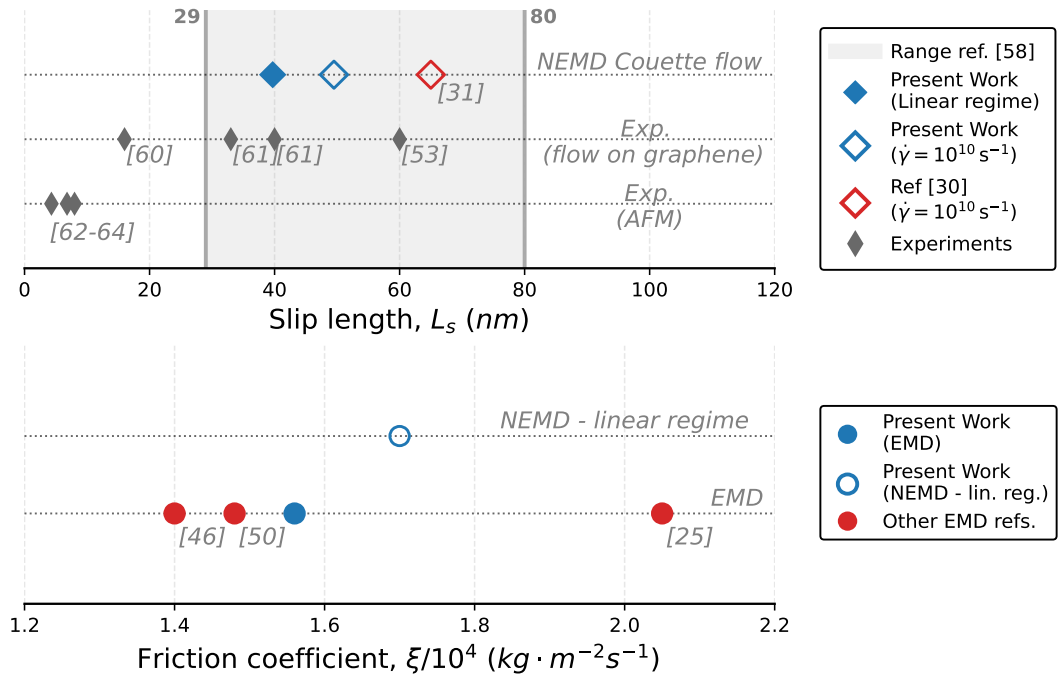

**Figure 6:** NEMD/TTCF and EMD results for the slip length (top panel) and the friction coefficient (bottom panel) reported in this study compared with those taken as a reference from the literature.

the validity of the present methodology.

# Acknowledgements

The authors thank the Australian Research Council for a grant obtained through the Discovery Projects Scheme (Grant No. DP200100422) and the Royal Society for support via International Exchanges (Grant No. IES/R3/170/233). J.P.E. was supported by the Royal Academy of Engineering (RAEng) through their Research Fellowships scheme. D.D. was supported through a Shell/RAEng Research Chair in Complex Engineering Interfaces. The authors acknowledge the Swinburne OzSTAR Supercomputing facility, which is located on the traditional lands of the Wurundjeri people, and the Imperial College London Research Computing Service (DOI:10.14469/hpc/223) for providing computational resources for this work.

# Data Availability

The data supporting the findings of this work are available within the article and also in the supplementary material.